\newcommand{\bea}{\begin{eqnarray}}
\newcommand{\eea}{\end{eqnarray}}
\documentclass[twocolumn,pra,preprintnumbers,amsmath,amssymb,superscriptaddress]{revtex4}
\begin{document}

\preprint{CERN-PH-TH-2010-140}
\preprint{RIKEN-MP-3}

\title{Nucleon Statistics in Holographic QCD : \\
Aharonov-Bohm Effect in a Matrix Model}

\author{Koji {\sc Hashimoto}}\email[]{koji(at)riken.jp}
\affiliation{
{\it Mathematical Physics Lab., RIKEN Nishina Center, Saitama 351-0198,
Japan }}

\author{Norihiro {\sc Iizuka}}\email[]{norihiro.iizuka(at)cern.ch}
\affiliation{
{\it Theory Division, CERN, CH-1211 Geneva 23, Switzerland}}

\begin{abstract}
We show that the Aharonov-Bohm effect in 
the nuclear matrix model \cite{Hashimoto:2010je}
derives the statistical nature of nucleons in holographic QCD.
For $N_c$= odd (even), the nucleon is shown to be a fermion (boson).
\end{abstract}

\maketitle


The statistics of baryons depends on the number of colors in QCD; 
in particular for large $N_c$ QCD, as the baryons are bound states of
$N_c$ quarks, they  
are fermions for odd $N_c$, while bosons  for even $N_c$. The nuclear
matrix model \cite{Hashimoto:2010je} derived in holographic QCD
offers a simple effective description of multi-baryon systems, where 
we can compute baryon spectra, short-distance nuclear forces, 
and even three-body nuclear forces \cite{Hashimoto:2010ue}. 
However, since the nuclear matrix model has only bosonic variables,  
it is natural to ask how the fermionic nature of baryons comes out from
the matrix model.  In chiral soliton models, this question 
was answered from the properties of Wess-Zumino term \cite{Witten:1983tx}.

To identify the statistics (fermionic/bosonic) of nucleons in the
nuclear matrix model, we consider a $2\pi$ rotation in the target space
of the matrix model. The target space index is carried by $X^M$ and
$w_{\dot{\alpha} i}$. The effect on $X^M$ is trivial, since $X$ decouples
from the system in the matrix model for a single baryon
($k=1$) once the ADHM constraint is solved. However, 
since we have a nontrivial gauge field
$A_0$, there is a
nontrivial effect on the $w_{\dot{\alpha} i}$ sector. 
In fact, this gauge field $A_0$ turns out to be responsible for
the statistics of the baryons, as we will see.

In a pion effective lagrangian, 
it is known that the Wess-Zumino term 
is essential for showing the nucleon statistics, in
the picture of solitonic nucleon of the system \cite{Witten:1983tx}. 
Now, in holographic QCD,
this Wess-Zumino
term is known to be from the 4-form Ramond-Ramond flux in the
gravity background in the D4-D8 model of holographic QCD
\cite{Sakai:2004cn}. 
In the nuclear matrix model \cite{Hashimoto:2010je}, the Ramond-Ramond
flux generates a Chern-Simons term in 1 dimension, which is just a term
consisting of a single gauge field $A_0$. The $w_{\dot{\alpha} i}$ field
is charged under the gauge symmetry, so it is 
natural to expect that the gauge dynamics in this 1 dimension with the
Chern-Simons term gives the nucleon statistics. 

In the nuclear matrix model, the terms including the fundamental field 
$w_{\dot{\alpha} i}$, except for the ADHM potential terms and the mass
term, are 
\begin{eqnarray}
 S = \frac{\lambda N_c M_{\rm KK}}{54\pi} \int \! dt
\; 
D_0 \bar{w}_i^{\dot{\alpha}} D_0 w_{\dot{\alpha}i}
+ N_c \int \! dt \; A_0 \, .
\end{eqnarray}
$\dot{\alpha}$ is a spinor index which is for $SU(2)\simeq SO(3)$ spatial
rotation in the target space. $i=1,\cdots, N_f$ is a flavor index. This
is a one-dimensional gauge theory whose gauge field is $A_0$. The
covariant derivative is defined as 
$ D_0 w_{\dot{\alpha}i} \equiv \partial_0 w_{\dot{\alpha}i} 
- i w_{\dot{\alpha}i}A_0.$
Note that $A_0$ is a gauge field for $U(k)$ gauge symmetry of the matrix
model, so, for $k=1$ (single baryon), 
$A_0$ does not carry any non-Abelian index.

Let us make a spatial rotation, for example along the $x^3$ axis, by
an angle $2\pi$. We look at how a wave function of a baryon transforms
under this rotation, and if it acquires a phase $n\pi$ with an 
odd integer  $n$, {\it i.e.} it
changes a sign, then the state is determined to be a fermion. 

Since $w_{\dot{\alpha} i}$ carries a spinor index of the target space,
it is obvious that 
the spatial rotation acts for the case of the rotation around the $x^3$
axis with an angle $\theta$, as
\begin{eqnarray}
 w_{\dot{\alpha}i} \to U_{\dot{\alpha}}^{\;\dot{\beta}}
w_{\dot{\beta}i}\, , \quad 
 U = \exp \left[ i \frac{\theta}{2} \tau^3\right]\, .
\label{spatialrot}
\end{eqnarray}
Here $\tau^3$ is the third component of Pauli matrices.
Our spatial rotation by $2\pi$ means that the angle $\theta$ moves in
the period $0 \leq \theta \leq 2\pi$.

As shown in \cite{Hashimoto:2010je}, the vacuum of the matrix model for
$k=1$ is quite simple, 
\begin{eqnarray}
 w = \left(
\begin{array}{cc}
\rho_0 & 0 \\
0 & \rho_0
\end{array}
\right)\, .
\label{vac}
\end{eqnarray}
After minimizing the hamiltonian, we obtain a certain nonzero value for
this $\rho_0$. So the spatial rotation (\ref{spatialrot}) corresponds to
a certain path in the target space of $w_{\dot{\alpha} i}$. In the
following, we would like to
compute an Aharonov-Bohm phase with this path. For that, it is
inconvenient that two nonzero entries in (\ref{vac}) moves
simultaneously. So, we combine a gauge transformation $\exp[-i\theta/2]$
together with the spatial rotation (\ref{spatialrot}),
so that we find a path 
\begin{eqnarray}
 w_{\dot{\alpha}i} \to U_{\dot{\alpha}}^{\;\dot{\beta}}
w_{\dot{\beta}i}\, , \quad 
 U = \left(
\begin{array}{cc}
1 & 0 \\ 0 & e^{-i\theta}
\end{array}\right) \, .
\label{path}
\end{eqnarray}
With this, we find that only the lower-right 
corner of $w_{\dot{\alpha} i}$ in (\ref{vac})
rotates. Indeed, the same change of the parameterization of the
path was used in \cite{Witten:1983tx} for the soliton in the 
pion effective field theory.


We are interested in a phase change of a baryon wave function. The
argument of the wave function is the moduli of this matrix model, and 
it is a part of $w_{\dot{\alpha} i}$ configuration space. If we think of
the path of $w_{\dot{\alpha} i}$ 
defined by (\ref{path}), then the phase of the wave function
of our concern is in fact an
Aharonov-Bohm (AB) phase, for the path (\ref{path}), as if we regard the
lower-right entry of the matrix 
field $w_{\dot{\alpha} i}$ as a position of a charged particle.

Let us write down the lagrangian for this charged particle, to compute
the AB phase. Writing the lower-right component of $w_{\dot{\alpha} i}$
as $w_{\dot{\alpha}=2, i=2} = u + iv$
where $u$ and $v$ are real, then the relevant part of the matrix model
is
\begin{eqnarray}
 S = \frac{\lambda N_c M_{\rm KK}}{54\pi}
\int \! dt \; |\partial_t (u + iv) - i A_0 (u+iv)|^2\, .
\label{uv}
\end{eqnarray}
It was shown in \cite{Hashimoto:2010je} that solving the equation
of motion for $A_0$, gives 
$A_0 = -27\pi/2 \lambda M_{\rm KK} \rho_0^2$,
which is a real constant.
Then the action (\ref{uv}) can be rewritten with conjugate momenta in
real coordinates as 
\begin{eqnarray}
 S = \frac{1}{2M} 
\int \! dt \; \left[
(P_u + A_0 v M)^2
+ (P_v - A_0 uM)^2
\right] \, .
\label{uv3}
\end{eqnarray}
Here we have defined the ``mass'' $M$ of the hypothetical particle
moving in the $u$-$v$ space as 
$M =  \lambda N_c M_{\rm KK}/27\pi$.
The expression shows that the particle is in a minimally-coupled 
gauge potential in the $u$-$v$ space, defined
by 
\begin{eqnarray}
 \widetilde{A}_u \equiv -A_0Mv = \frac{N_c}{2\rho_0^2}v\, , \quad
 \widetilde{A}_v \equiv A_0Mu = -\frac{N_c}{2\rho_0^2}u\, .
\label{gauge}
\end{eqnarray}
The magnetic flux made by this gauge potential is constant.

The path of this hypothetical charged particle is given by (\ref{path}), 
which is 
\begin{eqnarray}
 u+iv = \rho_0 e^{-i\theta} \, \quad (0\leq \theta \leq 2\pi)\, 
\end{eqnarray}
so the circle encloses the area $\pi \rho_0^2$, in a counter-clockwise
way. The AB phase $\Phi$ is 
given by an integration of the gauge potential (\ref{gauge}) along
this path,
\begin{eqnarray}
\Phi = - \rho_0 \oint\! \widetilde{A}_\theta d\theta = N_c \pi \, .
\label{AB}
\end{eqnarray}
In the last equality, we have used (\ref{gauge}) in a polar coordinate,
$\widetilde{A}_\theta = - N_c/2\rho_0$.
The negative sign is from the orientation of the path.

This AB phase means that, when $N_c$ is odd, the spatial rotation by
the angle $2\pi$ results in a sign $(-1)$ multiplied to the baryon wave
function. Therefore, when $N_c$ is odd (even), 
the baryon is a fermion (boson).

It is intriguing that a simple mechanism, the AB phase, is encoded in
the nuclear matrix model naturally to ensure the baryon statistics in
holographic QCD.

\vspace{5pt}
\noindent
{\bf --- Acknowledgment.}
We would like to thank Y.~Tachikawa for bringing us a 
question, which motivated the present work.
K.H.~is partly supported by
the Japan Ministry of Education, Culture, Sports, Science and
Technology.

\end{document}